\documentclass[abbrv,aps,prb,twocolumn,showpacs,preprintnumbers,amsmath,amssymb,
superscriptaddress,nofootinbib]{revtex4}

\usepackage{graphicx}

\begin{document}
\DeclareGraphicsExtensions{.ps,.jpg}


\input epsf

\title{Metal-insulator transition and nature of the gap in NiS$_{2-x}$Se$_x$}
\author{J. Kune\v{s}}
\affiliation{Theoretical Physics III, Center for Electronic Correlations and Magnetism, Institute of Physics,
University of Augsburg, Augsburg
86135, Germany}
\email{jan.kunes@physik.uni-augsburg.de}
\affiliation{Institute of Physics,
Academy of Sciences of the Czech Republic, Cukrovarnick\'a 10,
162 53 Praha 6, Czech Republic}
\author{L. Baldassarre} 
\affiliation{Experimental Physics II, Institute of Physics,
University of Augsburg, Augsburg
86135, Germany}
\author{B. Sch\"achner} 
\affiliation{Experimental Physics II, Institute of Physics,
University of Augsburg, Augsburg
86135, Germany}
\author{K. Rabia}
\affiliation{Experimental Physics II, Institute of Physics,
University of Augsburg, Augsburg
86135, Germany}
\author{C. A. Kuntscher} 
\affiliation{Experimental Physics II, Institute of Physics,
University of Augsburg, Augsburg
86135, Germany}
\author{Dm.~M. Korotin}
\affiliation{Institute of Metal Physics, Russian Academy of
Sciences-Ural Division, 620041 Yekaterinburg GSP-170, Russia}
\author{V.~I. Anisimov}
\affiliation{Institute of Metal Physics, Russian Academy of
Sciences-Ural Division, 620041 Yekaterinburg GSP-170, Russia}
\author{J.~A. McLeod}
\affiliation{Department of Physics and Engineering Physics, University of Saskatchewan,
116 Science Place, Saskatoon, Saskatchewan, Canada S7N 5E2}
\author{E.~Z. Kurmaev}
\affiliation{Institute of Metal Physics, Russian Academy of
Sciences-Ural Division, 620041 Yekaterinburg GSP-170, Russia}
\author{A. Moewes}
\affiliation{Institute of Metal Physics, Russian Academy of
Sciences-Ural Division, 620041 Yekaterinburg GSP-170, Russia}
\date{\today}

\begin{abstract}
The origin of the gap in NiS$_2$ as well as the pressure and doping induced metal-insulator
transition in the NiS$_{2-x}$Se$_x$ solid solutions are investigated both theoretically by 
the first-principles bandstructures combined with the dynamical mean-field approximation
for the electronic correlations and experimentally by means of infrared
and X-ray absorption spectroscopy. The bonding--anti-bonding splitting
in the S-S (Se-Se) dimer is identified as the main parameter controlling
the size of the charge gap. The implications for the
metal-insulator transition driven by pressure
and Se doping are discussed. 
\end{abstract}

\pacs{71.30.+h, 62.50.-p, 78.30.-j}

\maketitle
\section{Introduction}
The metal-insulator transition (MIT) due to electronic correlations has been subject
of intense research for several decades. The NiS$_{2-x}$Se$_x$ series provided
an important model system exhibiting a MIT controlled by varying the Se content $x$, temperature $T$ or pressure $P$ \cite{kwi80,yao96,matsu00,miya00,tak07}.
Particularly interesting is the similarity of 
the $x-T$ \cite{yao96,matsu00} and $P-T$ phase diagrams \cite{tak07} to that of the Hubbard model in the infinite dimension limit \cite{rmp},
consisting in the presence of the phase transition between paramagnetic metal and paramagnetic insulator at intermediate temperatures, existence of the
high-temperature crossover regime, and the possibility of driving a metal insulating, in a certain range of $x$, by increasing the temperature.
Consequently, the MIT has been commonly attributed to broadening of the Ni-$d$ band \cite{fuj96,matsu96,sar98,matsu00,miya00,tak07}.
Despite a large volume of available experimental data, the microscopic 
origin of the MIT in NiS$_{2-x}$Se$_x$ is poorly understood and a satisfactory material-specific theory is missing.

Ni$X$$_2$ ($X$=S,Se) can be viewed as NiO with the O atom replaced by an $X$$_2$ dimer. 
Strong hybridization between the $X-p$ orbitals pointing along the $X-X$ dimer, $p_{\sigma}$ orbitals, 
leads to a formation of split bonding and anti-bonding bands (the latter ones are referred to as $p_{\sigma}^*$), which accommodate two holes 
leading to an $X$$_2^{-2}$ valence state.
Numerous photoemission (PES) studies \cite{pes_eastman,folk87,fuj96} complemented by cluster calculations \cite{fuj96} revealed a similarity 
between the Ni$X$$_2$ and NiO valence band (VB) spectra, supporting this analogy.
The bandstructure calculations in the local density approximation (LDA) \cite{bull82,matsu96,per08} rendered NiS$_2$ a metal 
with a partially filled $e_g$ band and empty $p_{\sigma}^*$ orbitals. By analogy 
to NiO it was speculated that local $d-d$ correlations open a gap between
the S-$p$ band and the upper Hubbard band of Ni-$d$ $e_g$ character \cite{matsu96}, leading
to the classification of NiS$_2$ as a charge-transfer (CT) insulator in the Zaanen-Sawatzky-Allen (ZSA) scheme \cite{zaa85}, 
although in the original ZSA paper NiS$_2$ was
classified as '$p$-type metal were it not for the gap between the S $p_{\sigma}^*$ and the rest
of the $p$ band'. 

We have combined theoretical calculations with new experimental data in order to address the following questions: i) Why is NiS$_2$ an insulator and NiSe$_2$ a metal?,
ii) How do Ni$X$$_2$ respond to external pressure?, iii) Are the effects of pressure and varying Se content equivalent?,
iv) What is the mechanism of the MIT in NiS$_{2-x}$Se$_x$? 
The electronic structure of Ni$X$$_2$ is studied numerically,
using a combination of the {\it ab initio} bandstructure and the approximation of 
the dynamical mean-field theory (LDA+DMFT) \cite{rmp,ldadmftb}.
The infrared (IR) reflectivity is measured on samples with various Se concentrations under applied pressure
to investigate the evolution of the optical gap and the spectral weight transfer.
The x-ray absorption spectroscopy (XAS) at S K-edge is used to elucidate the orbital character
of features in the single-particle spectrum of NiS$_2$.

\section{Computational method}
Our computations proceed as follows. First, we perform paramagnetic LDA calculations \cite{wien2k} 
and minimize the total energy to find the equilibrium S(Se) positions. 
Next, we represent the one-particle Hilbert space of the hybridized Ni-$d$ and S(Se)-$p$ bands in Wannier basis 
(44 orbitals per unit cell) and calculate the parameters of the on-site $d-d$ interaction with the constraint LDA approach \cite{kor08,esp}. 
We found only moderately different $U_{\text{NiS}_2}=5$~eV and $U_{\text{NiSe}_2}=4.7$~eV and
used $J$ of 1~eV throughout the study. Finally, we construct a multi-band Hubbard Hamiltonian
\begin{equation}
\label{eq:ham}
\hat{H}=\sum_k\hat{\mathbf a}^{\dagger}_k{\mathbf H_k}
\hat{\mathbf a}_k-\varepsilon_{dc}(n_d)\hat{N}_d \\
+\sum_{i\in \text{Ni}} \hat{\mathbf n}^d_i{\mathbf U}^{dd}\hat{\mathbf n}^d_i,
\end{equation}
and compute the single-particle spectra using the DMFT approximation.
Here, $\mathbf H_k$ is the LDA Hamiltonian, 44$\times$44 matrices
(four NiX$_2$ units) in the basis of Ni-$d$ and X-$p$ Wannier orbitals,
on a uniform $k$-mesh in the first Brillouin zone \cite{kor08}. The
second term is the double-counting correction amounting to a constant
shift applied to Ni-$d$ site energies \cite{kunes-fe2o3}. The last term is the two-particle
interaction at the Ni sites in the density-density approximation.
The DMFT equations are solved on the Matsubara contour. The continuous time quantum Monte-Carlo
impurity solver \cite{ctqmc} is employed to solve the auxiliary multi-orbital impurity problem. 
Single-particle spectral densities are obtained by analytic continuation of Monte-Carlo
data using the maximum entropy method \cite{maxent}.
\begin{figure}
\includegraphics[height=0.9\columnwidth,clip]{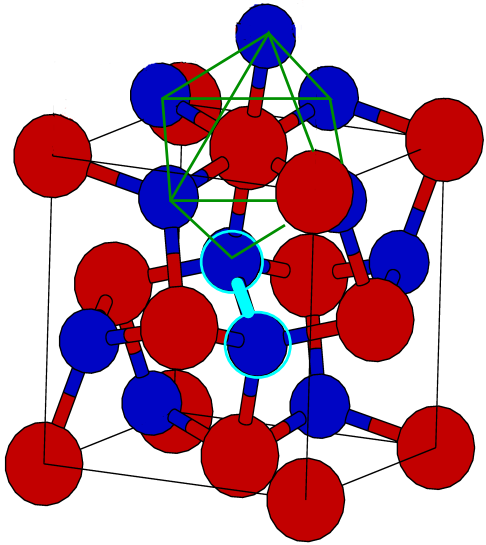}
\caption{\label{fig:str}(color online) The crystal structure of NiS$_2$: Ni atoms red (bright), S atoms blue (dark).
The S-S dimer is highlighted in the center of the figure.}
\end{figure}

\section{Experimental details}
Single crystals, grown by vapor transport technique, were slightly polished in order to obtain a clean, mirror-like surface. Near-normal incidence reflectivity was measured with a cassegrainian-based IR microscope coupled to a Michelson interferometer. 
The NiS$_{2-x}$Se$_{x}$ series (x=0, 0.3, 0.5, 0.7) was studied at ambient conditions,

A clamp diamond anvil cell (Diacell cryoDAC-Mega) equipped with type IIa diamonds was employed to perform pressure dependent studies. 
A small piece of NiS$_2$ (NiS$_{1.7}$Se$_{0.3}$) was cut and placed together with CsI (as hydrostatic medium) inside a 400 $\mu$m hole drilled in a 90 $\mu$m thick CuBe gasket. The pressure was evaluated in situ, by the standard ruby-fluorescence technique.\cite{Mao}
Measurements were carried out partly at the University of Augsburg and partly at the IR1 beamline in ANKA, so to exploit the high brilliance of synchrotron radiation in the far-IR range. We recall that due to the small size of the samples inside the diamond anvil cell measurements as a function of pressure were performed for $\omega>$250 cm$^{-1}$.
The measurement procedure, as well as the data processing to extract the optical conductivity $\sigma_1(\omega)$ are carefully described in Refs. \onlinecite{Pashkin,Baldassarre}. 

The sulfur 1s X-ray absorption spectra (XAS) spectra which provides information about vacant S 3p-states were 
measured at the soft X-ray microcharacterization beamline (SXRMB) at the Canadian Light Source. 
The spectra were measured in total electron yield (TEY) mode. Additional measurements were taken in total 
fluorescence (TFY) mode using a channel-plate detector, and partial fluorescence (PFY) mode using a silicon 
drift detector with a resolution of about 100 eV. Both the PFY and TFY measurements showed the same features as the 
TEY measurements, indicating that the surface of the samples was clean. The TEY measurements were used for the analysis 
because they had a better signal-to-noise ratio than the TFY or PFY measurements, and did not suffer 
from the self-absorption effects present in the fluorescence measurements. 
An Si(111) crystal was used in the monochromator, and the experimental resolution (E/E) was about 10000.

\section{Results and discussion}
We start the exposition of our results with a brief review of the non-interacting LDA bandstructures,
which agree with those published previously \cite{bull82,matsu96,per08}. The Ni-$d$ orbitals
form narrow $t_{2g}$ and broader, Ni-X anti-bonding, $e_g^{\sigma}$ bands due to octahedral coordination of Ni with the chalcogen X.
The deviations from an exact octahedral symmetry at the Ni sites lead to further splitting of the $t_{2g}$ bands
into an $e_g^{\pi}$ doublet and an $a_{1g}$ singlet. This splitting plays no discernible role in our investigation.
The characteristic feature of the chalcogen bands is a large bonding--anti-bonding splitting of the $p$-orbitals
pointing along the X-X dimer (Fig.~\ref{fig:str}). Due to the S-S-Ni angle being close to the right angle ($\approx$104 deg in NiS$_2$)
the dimer states hybridize only weakly with the Ni $e_g^{\sigma}$ states.

The Ni-X and X-X dimer distances as function of the lattice constant were obtained
by minimization of the LDA total energies (see Fig.~\ref{fig:str}). At ambient pressure the calculated NiS$_2$ bond lengths agree well with experimental
observations. Compression leads to shortening of the Ni-S distance while the S-S dimer behaves as a rigid object.
The situation is less clear in NiSe$_2$. While the calculated ambient pressure Ni-Se bond length reproduces the 
experimental value, the length of the Se-Se dimer is substantially overestimated. Under pressure our calculations
yield an initial increase of the Se-Se indicating that the Ni-Se bond becomes stronger relative to the Se-Se bond and suggesting that
Se-Se dimer cannot be viewed as a molecular unit. We will discuss later on how the behavior of the bond lengths affects the electronic properties.

\subsection{Single-particle spectra}

\subsubsection{NiS$_2$}

\begin{figure}
\includegraphics[height=0.9\columnwidth,angle=270,clip]{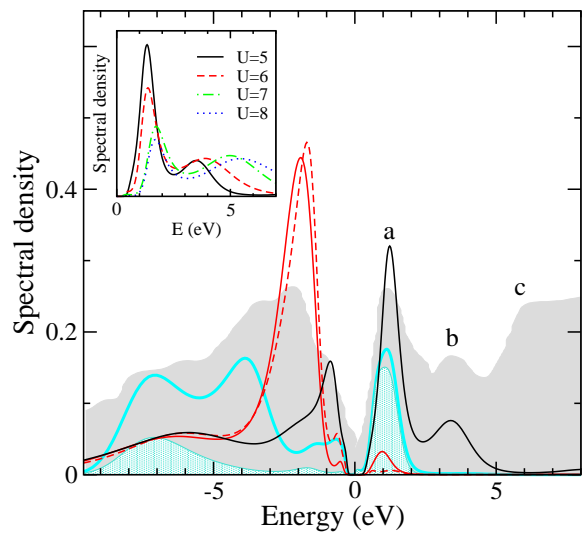}
\caption{\label{fig:nis2}(color online) The calculated ($U$=5~eV, $T$=580~K) orbitally resolved spectra of NiS$_2$
[Ni-$d$:$e_g^{\sigma}$ (black) and $t_{2g}$=$e_g^{\pi}$ (red) + $a_{1g}$ (red - dashed);
S-$p$: total (blue, thick) and $p_{\sigma}$ (blue-shaded)] compared to
the experimental XPS+BIS (gray, shaded) \cite{folk87}. The $a$, $b$, and $c$ are defined in the text.
The inset shows the conduction band $e_g^{\sigma}$ spectra for $U$=5, 6, 7, and 8~eV.}
\end{figure}

In this section we present orbitally resolved spectral densities obtained at ambient pressure with the LDA+DMFT method.
In Fig.~\ref{fig:nis2} we show the NiS$_2$ spectra, which exhibit a small charge gap in
agreement with experimental observations. The valence band (VB) resembles that of NiO \cite{kunes-nio} 
with the characteristic distribution of the Ni-$d$ spectral weight between $d^7$ (around -6 eV) and 
$d^8\underline{L}$ excitations (at low binding energy), the result of a strong $p-e_g^{\sigma}$ hybridization.
Overall, a good correspondence between positions of the main features in the experimental x-ray photoemission 
spectra (XPS) and their theoretical counterparts can be stated. 
Particularly important for the further discussion is the high-energy shoulder at -7~eV, which is 
dominated by the S-$p$ density in agreement with the observation of Ref. \onlinecite{pes_eastman}.
The absence of an obvious gap in the experimental spectra shown in Fig.~\ref{fig:nis2} is due
to the experimental resolution.
\begin{figure}
\includegraphics[height=0.85\columnwidth,angle=270,clip]{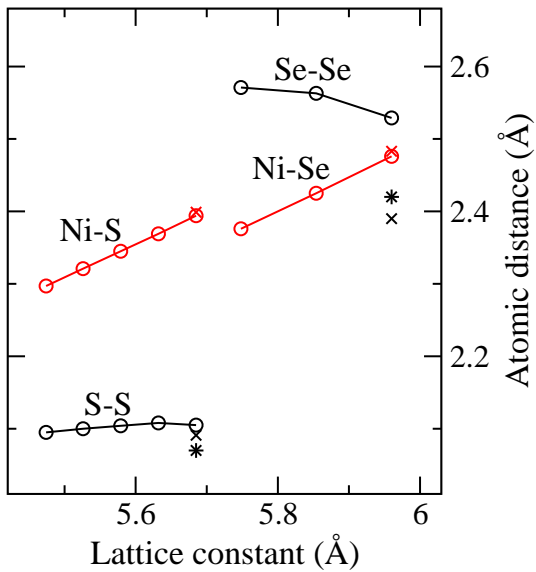}
\caption{\label{fig:dist}(color online) 
The calculated pressure dependence of the equilibrium Ni-X and X-X distances.
The isolated points mark the experimental values of Refs. \onlinecite{folk87} (star) and
\onlinecite{otero98} (cross).}
\end{figure}

The calculated conduction band (CB) spectrum consists of two peaks: ($a$) $p_{\sigma}^*$ band with a large 
resonant $e_g^{\sigma}$ contribution at 1~eV, and ($b$) the upper Hubbard band of pure $e_g^{\sigma}$ character
at 3.5~eV. The identification of $b$ with the upper Hubbard band is based on the observation of its shift
upon increasing the interaction strength (see the inset of Fig. \ref{fig:nis2}) and the absence of $p$ character. 
The experimental CB spectrum, obtained with the Bremsstrahlung
isochromat spectroscopy (BIS) of Ref. \onlinecite{folk87}, exhibits three distinct features: a sharp low-energy peak at 1 eV,
a high-energy peak at 3.5~eV and a broad band ($c$) starting around 5.5~eV. 
The 5.5~eV band is not spanned by the Hamiltonian (\ref{eq:ham}) and thus the corresponding
contribution is missing in Fig. \ref{fig:nis2}.
The orbital character of the other two peaks is of key importance for understanding the nature of the 
charge gap. The question can be formulated simply as:
'is the upper Hubbard band located below or above the $p_{\sigma}^*$ band'?  
The authors of Ref. \onlinecite{folk87} interpreted the 1~eV as Ni-$d$ upper Hubbard band and the 3.5~eV peak as
S $p_{\sigma}^*$ band, which is a clear contradiction to our result.

\begin{figure}
\includegraphics[height=0.9\columnwidth,angle=270,clip]{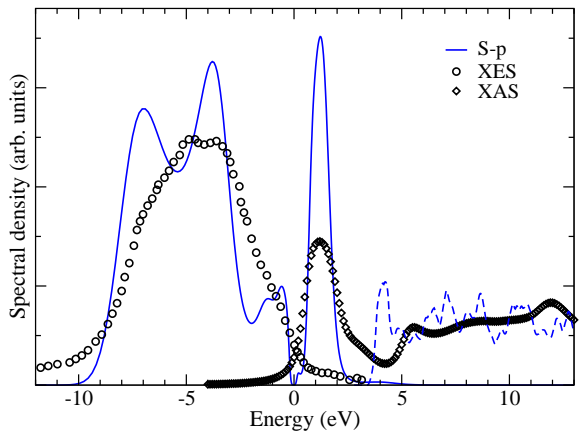}
\caption{\label{fig:xas} The calculated S-$p$ spectral density obtained from LDA+DMFT (full line), augmented
with the S-$p$ contribution from the high energy LDA bands (dashed line), compared to experimental XES spectral of Ref. \onlinecite{sug76} (circles)
and XAS (diamonds) spectra of this work.}
\end{figure}
In order to resolve this issue we have measured the XAS at the K-edge of sulfur, which can be directly
compared to the calculated S-$p$ spectral density as shown in Fig. \ref{fig:xas}. The theoretical 
high energy spectrum was obtained by projecting out the S-$p$ contribution from the LDA bands not included
in the Hamiltonian (\ref{eq:ham}). 
The identification of peak $a$ with $p_{\sigma}^*$ is corroborated by its position with respect to the 
X-ray emission (XES) spectrum as well as by its distance from the high energy (5~eV) XAS band. 
Although the calculated onset of the high energy band is misplaced by about 1~eV with respect to 
the experiment, the overall agreement is very good. We point out that the experimental XAS $a$ and $c$ peaks
match their BIS counterparts perfectly.

\subsubsection{NiSe$_2$}
\begin{figure}
\includegraphics[height=0.9\columnwidth,angle=270,clip]{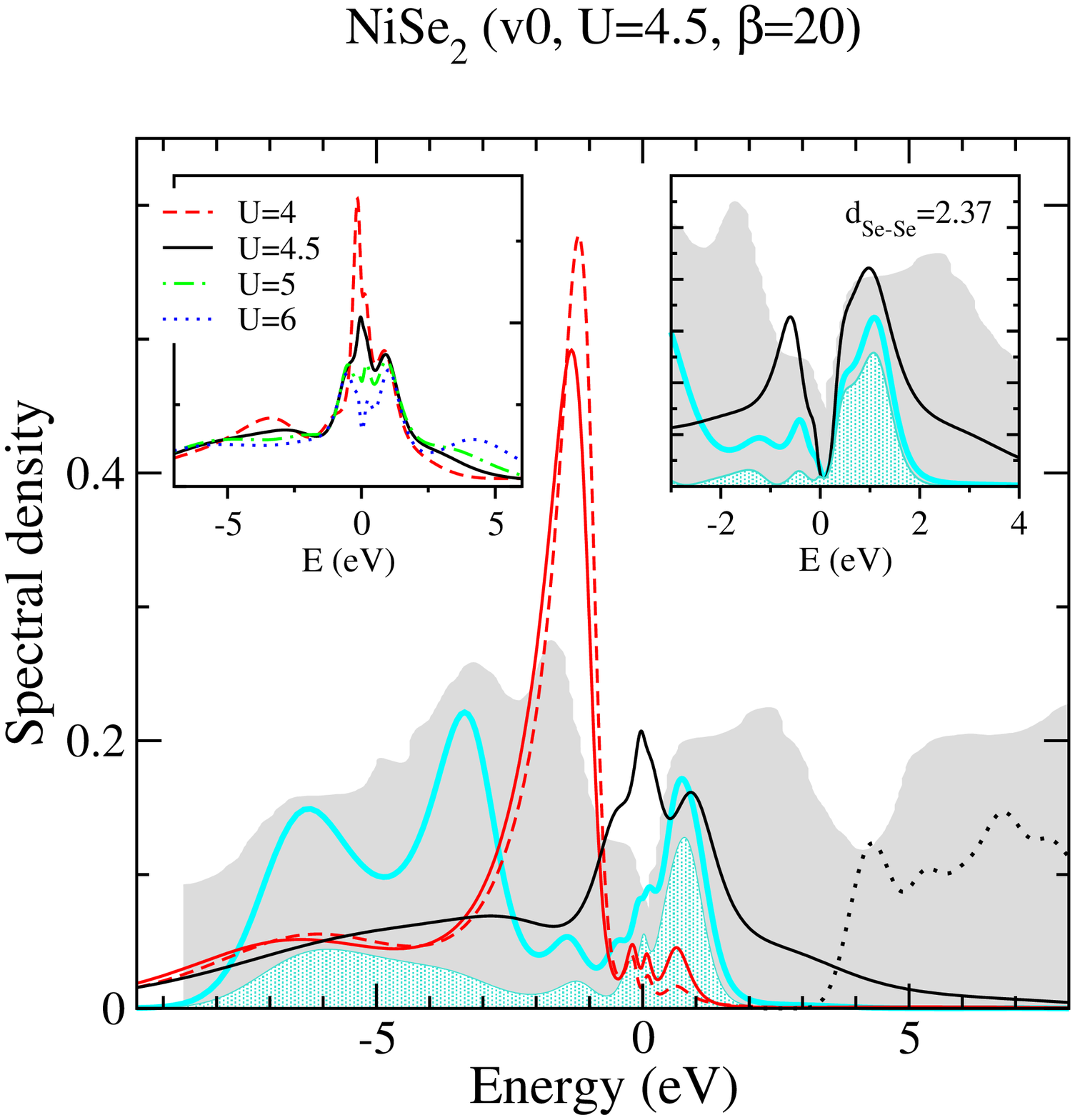}
\caption{\label{fig:nise2}(color online) The calculated ($U$=4.5~eV, $T$=580~K) orbital resolved spectra of NiSe$_2$
compared to the experimental XPS+BIS \cite{folk87} spectra. (the same notation as in Fig.~\ref{fig:nis2})
The position of the calculated high band is marked with the dotted line.
The left inset shows the $e_g^{\sigma}$ densities for various interaction strengths $U$ (eV).
In the right inset the effect of reducing the Se-Se distance ($\AA$) on the spectra ($U$=4.5~eV)
is shown.}
\end{figure}

The LDA+DMFT calculations yield NiSe$_2$ metallic. Its low-energy
spectrum is rather sensitive to changes of $U$, as shown in the left inset of Fig.~\ref{fig:nise2}. 
For $U<$5~eV, a temperature-dependent peak appears at the chemical potential. Nevertheless,
the metallic state is robust and the spectrum
remains gapless up to the largest studied value of $U$=8~eV (not shown).

The VB spectra NiSe$_2$ (Fig.~\ref{fig:nise2}) overall resemble the NiS$_2$ ones
with some quantitative differences, such as the shift of the high-energy shoulder to -6.5~eV, also visible
in the experimental data.
As in NiS$_2$ we find the $p_{\sigma}^*$ band at the bottom of the CB
manifold. Comparison of the calculated CB spectrum with the BIS data is less satisfactory
than in the case of NiS$_2$. In particular, the experimental 2~eV peak is not found in the 
calculation. 

Bearing in mind that the Se-Se dimer length of 2.52~\AA\ is overestimated, we have performed 
a reference calculation with a reduced dimer length of 2.37 \AA\ (just below the experimental values).
As shown in the right inset of Fig.\ref{fig:nise2},
an increase of the dimer bonding -- anti-bonding splitting leads to a strong suppression of the spectral
density at the lowest energies (see the right inset of Fig.~\ref{fig:nise2}), but is not sufficient
to open a charge gap. 
Since the bonding--anti-bonding splitting between the VB bottom and the $p_{\sigma}^*$ peak is essentially
a molecular property of the Se-Se dimer, it does not depend on the value of $U$.



\subsection{Origin of the charge gap}
\begin{figure}
\includegraphics[height=\columnwidth,angle=270,clip]{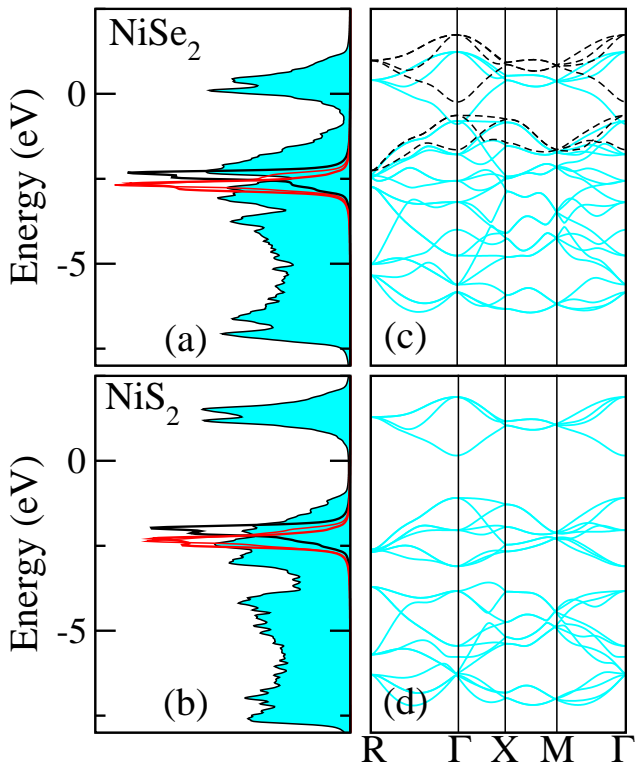}
\caption{\label{fig:trick}(color online) The bare S(Se)-$p$ spectral densities and bandstructures.
The spectral densities in the left panel [(a),(b)] are obtained from a Wannier function Hamiltonian,
while the bandstructures in the middle panel [(c),(d)] are based on a LAPW calculation with removed Ni-like
basis functions. Dashed bands for NiSe$_2$ correspond to a reduced Se-Se distance.}
\end{figure}

To analyze the behavior of the $p_{\sigma}^*$ band we have calculated the bare $p$ bandstructures,
obtained by decoupling the Ni-$d$ states.
This can be done conveniently by removing
the Ni-centered orbitals from the LDA basis set [Fig.~\ref{fig:trick}(c),(d)], 
or rigorously using the $p-p$ block of the Hamiltonian in Wannier basis [Fig.~\ref{fig:trick}(a),(b)]
with qualitatively similar results. The $p$-bands form a broad manifold bounded
by dimer bonding bands at the bottom and a distinct anti-bonding $p_{\sigma}^*$
band complex at the top, separated by a (pseudo)gap. 
The key difference between NiS$_2$ and NiSe$_2$
is the size of this gap separating the $p_{\sigma}^*$ band from the rest of the $p$-manifold
controlled by the length of the dimer.
The gap in NiS$_2$ is large enough to sustain the coupling to the correlated Ni-$d$ orbitals,
while the NiSe$_2$ gap is too small and the coupling to $d$ orbitals closes it completely.


\begin{figure}
\includegraphics[height=\columnwidth,angle=270,clip]{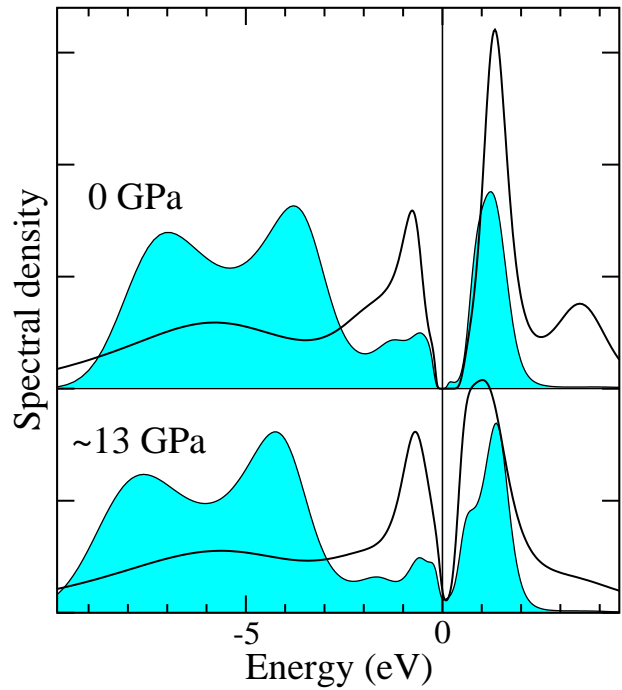}
\caption{\label{fig:press}(color online) The comparison of NiS$_2$ spectra in the insulating state at ambient pressure (upper panel)
and in the metallic state under compression (lower panel). The black line marks the $e_g^{\sigma}$, is shaded area (blue) is the 
total S $p$ contribution. The $t_{2g}$ spectra which show very little change with pressure are not shown.}
\end{figure}

Next, we consider the effect of applied pressure.
The LDA+DMFT calculations for pressures up to 13~GPa show the closing of the
charge gap (Fig. \ref{fig:press}). This is a result of the combination
of the rigidity of the S-S dimer and an overall band broadening.
As a consequence of the former the bonding -- anti-bonding splitting is
insensitive to pressure. On the other hand, the widths of the individual
$p$-bands, controlled by the reduced inter-dimer distances, increase with pressure
thus reducing the bare $p$ gap. Together with increasing $p-d$ hybridization this
leads to the metal-insulator transition.
Noteworthy the LDA results showed an elongation of the Se-Se dimer
at moderate pressures, suggesting that the pressure-induced closing 
of the charge gap proceeds faster in compounds with higher Se content.

\subsection{IR reflectivity}
\begin{figure}
\includegraphics[width=\columnwidth,angle=0,clip]{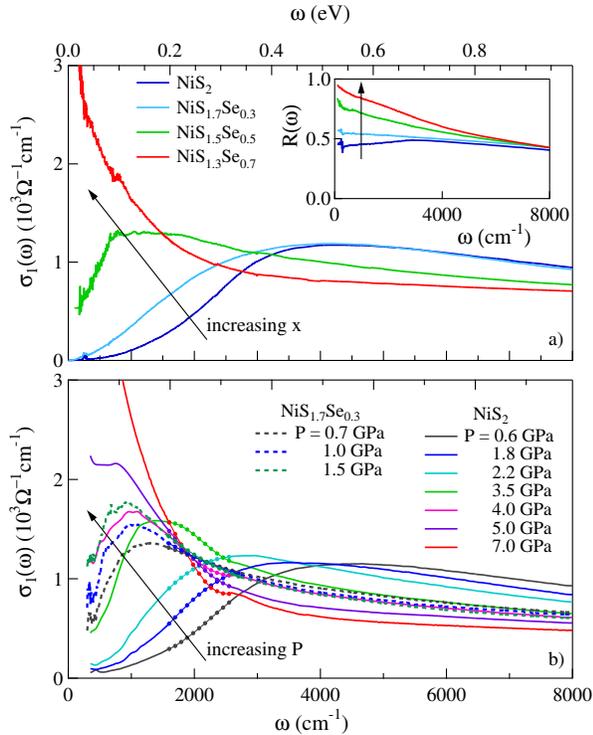}
\caption{\label{exp_data1}(color online) (a) Room-temperature optical conductivity spectra of NiS$_{2-x}$Se$_x$ ($x$= 0; 0.3; 0.5; 0.7, as indicated in figure) at ambient pressure. In the inset the reflectivity R($\omega$) is reported. (b) Optical conductivity of NiS$_2$ (solid) and NiS$_{1.7}$Se$_{0.3}$ (dashed) for increasing pressure.} 
\end{figure}

To verify the theoretical conclusions and to obtain further information about the electronic properties of NiS$_{2-x}$Se$_x$ 
we have measured the infrared properties of the NiS$_{2-x}$Se$_x$ series 
($x$=0; 0.3; 0.5; 0.7) at ambient conditions, shown in Fig.\ref{exp_data1}a). For NiS$_2$, R($\omega$) slightly increases with increasing frequency, showing a broad maximum in the mid-infrared  and two narrow phonon absorption in the far-IR. As Se is chemically substituted in the compound, R($\omega$) increases its absolute value at low frequency, partially shielding the phonon lines and eventually becoming metallic-like (\emph{i.e.} decreasing with increasing $\omega$) for $x$= 0.7.
The corresponding optical conductivities $\sigma_1(\omega)$ clearly show how the transition from an insulator (NiS$_2$) to a metal (NiS$_{1.3}$Se$_{0.7}$) takes place: A well-defined gap followed by an absorption band is found for NiS$_2$. As Se is introduced the absorption band moves towards lower energies, progressively closing the gap. 

A similar behavior is found as pressure is applied: as shown in Fig.\ref{exp_data1}b) it is again the red-shift of the absorption band that closes the gap. A closer inspection at the curve at 5 GPa for NiS$_2$ shows the coexistence of two distinct terms at low-frequency. As it was also shown in recent IR measurements  
\cite{per08}, our data can be fitted at low frequency with two distinct features: a Drude-like term and a Lorentz peak. 

\begin{figure}
\includegraphics[width=\columnwidth,angle=0,clip]{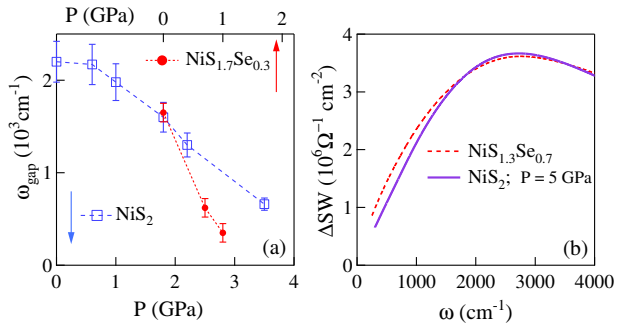}
\caption{\label{exp_data2}(color online) (a) The gap vs pressure for NiS$_2$ and NiS$_{1.7}$Se$_{0.3}$ (note the mutual horizontal offset)
3(c) SW transfer $\Delta$SW for NiS$_2$ at 5~GPa and for NiS$_{1.3}$Se$_{0.7}$ 
at ambient pressure.} 
\end{figure}

Our aim is to follow and compare the gap-closure by chemical alloying and external pressure.
We define the experimental optical gap as the frequency at half-height of the absorption band and report its values in Fig.\ref{exp_data2} for NiS$_2$ and NiS$_{1.7}$Se$_{0.3}$ under pressure. 
This definition of the gap leads to a similar pressure dependence as obtained by linearly extrapolating the steepest rise of $\sigma_1(\omega)$ \cite{katsufuji} or by considering the maximum of the absorption peak obtained with Drude-Lorentz fitting \cite{Wooten}. 
We remark that strong phonon absorptions of the diamond prevent to have (fully) reliable data between 1700-2500 cm$^{-1}$. To overcome this problem and unambiguously define the gap value, the data analysis was performed as follows: reflectivity data are fitted with a Drude-Lorentz model in order to extrapolate the missing parts of the spectra. The KK transformations are performed, followed by a simultaneous fitting of R and sigma. This process is carried out iteratively until the $\sigma_1(\omega)$ is stable with respect to  the extrapolation used in the KK transformations.

Comparing the conductivity spectra we find that the qualitatively similar effect of pressure on NiS$_2$ and S substitution by Se follows the scaling relation $P$[GPa]$\approx$$x$/0.15 proposed in Ref. \onlinecite{miya00}.
The pressure dependence of the optical gaps for pure NiS$_2$ and
NiS$_{1.7}$Se$_{0.3}$ is shown in Fig.\ref{exp_data2}a), with an horizontal off-set according to the scaling relation. Our experimental results show that in the Se-doped compound pressure is more efficient in closing the gap, in accord with the previous theoretical reasoning. 

Next we discuss the metallic phase. To quantify the 'metallicity' we use the spectral
 weight (SW) transfer relative to the ambient pressure NiS$_2$ data $\Delta$SW=SW$(x,P)$-SW($x$=0,$P$=0), where
SW$(\omega)$=$\int_0^\omega \sigma_1(\omega') d\omega'$. 
Recalling that at low frequency there is the coexistence of two terms we can infer that
with applied pressure a large part of the spectral weight is transferred to the mid-IR region first, while the 
appearance of Drude-like peak occurs upon further pressure increase [see Fig.~\ref{exp_data1}(b)].
Without making a quantitative connection we point out that a qualitatively similar behavior
is observed in the DMFT results: As pressure is applied to NiS$_2$, due to the rigidity of the S-S dimers the transition occurs mainly due to band broadening not corresponding at first instance to the formation of a Drude term. Instead, when the Se-Se distance is varied we expect a fast response of the low-energy electrodynamics (right inset of Fig.~\ref{fig:nise2}).
Following the scaling relation we compare NiS$_2$ at 5~GPa with
NiS$_{1.3}$Se$_{0.7}$ at ambient pressure in Fig.~\ref{exp_data2}b). Although a similar
amount of SW is transferred below 4000~cm$^{-1}$, NiS$_{1.3}$Se$_{0.7}$ exhibits a higher SW transfer at lower frequencies
($<$2000~cm$^{-1}$) pointing to a more intense Drude-like term, also supporting our
conclusion about the faster onset of metallicity with Se substitution. 

\section{Conclusions}
Using the LDA+DMFT approach we have reproduced the paramagnetic phases of NiS$_2$ (insulator) and NiSe$_2$ (metal)
and provide the following answers to the questions posed in the introduction:
i) The presence of the charge  gap in NiS$_2$ in contrast to NiSe$_2$ is a consequence of a larger gap in the bare $p$-bandstructure ($p$-gap)
of NiS$_2$ resulting from stronger bonding within the S-S dimer. ii) The pressure reduces the
$p$-gap by broadening the individual $p$-bands, while the Se substitution reduces the $p$-gap
due to smaller bonding--anti-bonding splitting.
iii) The MIT is controlled by varying the size of the $p$-gap.
We find both computationally and experimentally that the pressure driven MIT is accelerated by the Se substitution.
The conceptual picture of NiS$_{2-x}$Se$_x$ is provided by a periodic Anderson model with
a variable gap in the conduction electron bath rather than the charge-transfer insulator model.

\acknowledgements
We thank P. Werner for providing his quantum Monte-Carlo code, 
 D. Vollhardt for critical reading  of the manuscript and C. Schuster for useful discussions.
L.B., K.R., and C.A.K. thank B. Gasharova, Y.-L. Mathis, D. Moss, and M. S\"upfle for help at the IR beamline and
H. Takagi for kindly providing the samples.
We acknowledge the financial support of SFB 484 of the Deutsche Forschungsgemeinschaft (J.K.,L.B.,C.A.K.,K.R.), Fondazione della Riccia (L.B.) and 
Bayerische Forschungsstiftung (L.B.,R.K.), and Russian Foundation for Basic Research under Grant No. RFFI-07-02-00041, Dynasty Foundation, 
President of Russian Federation fund NSH 1941.2008.2 and Program of Russian Academy of Science Presidium 
``Quantum microphysics of condensed matter'' N7 (Dm.M.K, V.I.A.), beam time provided by ANKA Angstr\"omquelle Karlsruhe
and computertime provided by Leibniz Supercomputing Center Munich.
J.A.M., E.Z.K and A.M. acknowledge support of the Research Council of the President of the Russian Federation (Grant No. NSH-1929.2008.2), 
the Russian Science Foundation for Basic Research (Project No. 08-02-00148), 
the Natural Sciences and Engineering Research Council of Canada (NSERC), and the Canada Research Chair program.

\end{document}